\documentclass[twoside]{article}

\usepackage[accepted]{aistats2026}

\usepackage{amsmath,amssymb,bbm}
\usepackage{stfloats}
\usepackage{graphicx} 
\usepackage{url}

\usepackage[round]{natbib}

\begin{document}

\twocolumn[

\aistatstitle{On the Calibration of Isotonic Distributional Regression}

\vspace{-0.06cm}

\aistatsauthor{Tobias Biegert \And Johannes Resin \And Alexander Jordan \And Sebastian Lerch}

\aistatsaddress{
Karlsruhe Institute\\of Technology
\And Goethe University\\Frankfurt
\And Heidelberg Institute for\\Theoretical Studies
\And Marburg University
}
]

\begin{abstract}
    Isotonic distributional regression (IDR) has recently been proposed as a non-parametric technique for probabilistic forecasting via the estimation of conditional distributions under order restrictions, based on continuous outcomes from deterministic model outputs. 
    IDR has been widely used and multiple studies report promising performance and well-calibrated forecasts across application domains.
    We document a peculiar out-of-sample artifact of IDR, where probability integral transform (PIT) values are rational fractions with empirical frequencies resembling Thomae's function, an effect which thus far seems to have been overlooked.
    This phenomenon is demonstrated empirically using meteorological data and a simulation study.
    We further provide a theoretical explanation linking the spiking behavior to the discrete structure of the isotonic fits, and discuss how this artifact can be mitigated via smoothing approaches. 
\end{abstract}

\section{INTRODUCTION}

IDR \citep{henzi2021idr} estimates the conditional distribution of a real-valued outcome $Y$ given a predictor $X$ based on training data of the form $(x_i, y_i), i = 1,...,n$, under a stochastic order constraint. 
IDR assumes that the conditional distribution of the outcome given by the cumulative distribution function (CDF) $F_x(y) = P(Y \leq y | X = x)$ increases in stochastic order as $x$ increases in partial order, i.e., if $x \preceq x^\prime$ then $F_x(y) \geq F_{x^\prime}(y)$ for all $y \in \mathbbm{R}$. 

In the following, we will restrict our attention to IDR for totally-ordered univariate model outputs. This special case has been introduced as EasyUQ by \citet{walz2024easyuq}.
The appeal of IDR and EasyUQ lies in their in-sample optimality, conceptual simplicity, and lack of tuning parameters.
If $x_1 < \cdots < x_n$, the IDR solution 
\begin{equation}
\label{eq:IDR}
	\widehat F_{x_j}(y) = \min_{k=1,...,j} \max_{l=j,...,n} \frac{1}{l-k+1} \sum_{i=k}^{l} \mathbbm{1}\{y_i \leq y\},
\end{equation}
is optimal relative to a comprehensive class of proper scoring rules for probabilistic forecasts \citep{henzi2021idr}.

IDR and EasyUQ have been used in various studies across application domains, including weather \citep{henzi2021idr,SchulzLerch2022,ageet2023rainfall,rasheedasatheesh2023,walz2024africa,buelte2025uqDataDriven,rasheedasatheesh2025, Kalita2026} and energy \citep{gneiting2023solar,Lipiecki2024}. 
These studies report promising performance, and, in most cases, well-calibrated forecasts obtained via IDR.

It is standard practice to assess the calibration of distributional predictions via a PIT histogram, which should show approximately uniform PIT values \citep[e.g.,][]{Dawid1984, Diebold1998}.
In this note, we demonstrate an interesting artifact in the PIT values of IDR forecasts that appears  to have been overlooked, possibly because calibration diagnostics typically rely on PIT histograms with a relatively coarse binning (usually 10--20 bins).
When evaluating IDR forecasts, the PIT values show a peculiar pattern with pronounced spikes at simple rational fractions. Upon closer investigation, the empirical frequencies of the PIT values resemble \emph{Thomae's function} \citep{BartleSherbert2011}, which is defined by $f(x) = 0$ if $x$ is irrational, 
and $f(x) = 1/q$ if $x = p/q$ is a rational number written as a reduced fraction with $q > 0$. 

We provide theoretical background on calibration results for IDR (Section 2), demonstrate the PIT phenomenon in data and simulation examples (Section 3), and provide a theoretical explanation (Section 4). We conclude with a discussion of a remedy via smoothing (Section 5) and practical implications (Section 6). 

\section{BACKGROUND}

\begin{figure*}[t]
  \centering
  \includegraphics[width=\linewidth]{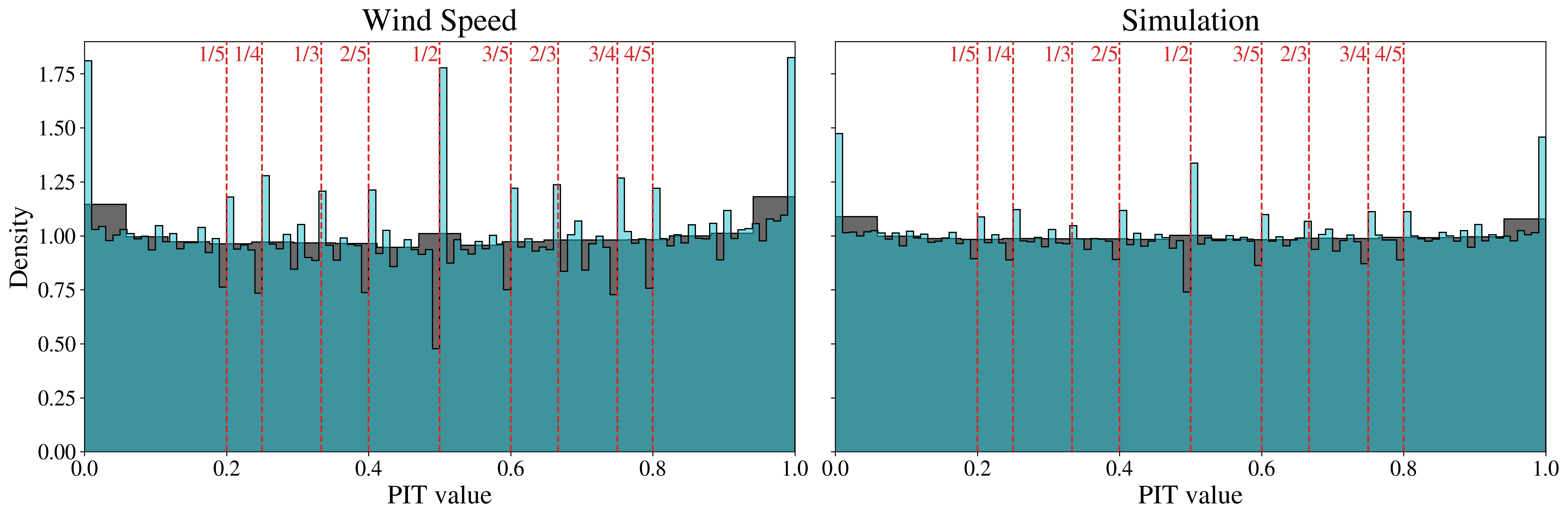}
  \caption{10-m wind speed (left) and simulation (right) out-of-sample PIT histograms for EasyUQ forecasts with coarse (17-bin; gray) and fine (100-bin; cyan) binning.
  Coarse binning largely masks systematic over-representation at simple rational PIT values, while finer binning reveals pronounced spikes.}
  \label{fig:pit_binning}
\end{figure*}

Following standard practice \citep{GneitingRanjan2013}, we assess \emph{probabilistic calibration} using the (randomized) PIT, 
\[
Z_F \;=\; \lim_{y\uparrow Y} F(y) \;+\; V\Bigl(F(Y)-\lim_{y\uparrow Y}F(y)\Bigr),
\]
where $F$ is a CDF-valued random quantity (i.e., a probabilistic forecast), $Y$ denotes the realized outcome, and $V\sim\mathrm{Unif}(0,1)$ is independent of $(F,Y)$.
If $F$ is continuous at $Y$, then $Z_F=F(Y)$. 
Probabilistic calibration of $F$ corresponds to $Z_F\sim\mathrm{Unif}(0,1)$, and is commonly assessed using PIT histograms.

The IDR solution \eqref{eq:IDR} consists of discrete distributions $\widehat F_{x_j}$ with possible jumps at the unique values $\widetilde y_1<\cdots<\widetilde y_k$ of the outcomes $y_1,\ldots,y_n$ in the training set. 
In-sample (i.e., with respect to the empirical distribution of the training data) the IDR solution enjoys good calibration properties. 
\citet[Theorem~2]{henzi2021idr} show that the IDR solution is threshold calibrated. While this does not imply probabilistic calibration \citep{gneiting2023calibration}, IDR satisfies a somewhat weaker version of probabilistic calibration \citep[see][Appendix D]{ArnoldZiegel2025}.

To obtain an out-of-sample predictive distribution for a new value of the predictor $x$, we follow \citet{henzi2021idr} and interpolate the IDR solution \eqref{eq:IDR} linearly, which produces another discrete CDF, $\widehat F_x$.
Below, we demonstrate that the discreteness of the predictive distributions induces systematic spikes in PIT values at simple rational fractions out-of-sample.  Thus, the IDR predictions fail to be probabilistically calibrated in a rather peculiar way.
This lack of calibration has thus far gone unnoticed, likely because coarse binnings in PIT histograms can mask the observed spikes.

\section{EMPIRICAL EVIDENCE}

We analyze global gridded 10-m wind speed forecasts obtained by applying IDR to point predictions from an AI-based weather model, which we evaluate out-of-sample against reanalysis data within the WeatherBench~2 framework \citep{rasp2024weatherbench}.
All forecasts use a 3-day lead time for 00/12\,UTC initialization times. 
We use data from the years 2018-2019 and 2021-2022 to fit IDR on each grid point individually and evaluate out-of-sample on 2020. The underlying grid has a resolution of $64\times 32$ with $732$ test time steps in 2020.

The left panel of Figure~\ref{fig:pit_binning} shows PIT histograms for the same set of out-of-sample PIT values using a fine and a coarse binning.
With the fine binning, PIT values are clearly clustered near simple rational fractions such as $1/2$, $1/3$, $2/3$, $1/4$, and $3/4$, with additional clusters near $0$ and $1$, and thus deviate from the expected uniform distribution under probabilistic calibration.
In contrast, when the same PIT values are displayed with a coarser binning, as is common in the literature, the spikes are largely hidden, and deviations from uniformity are much less apparent.

To illustrate that the observed spiking behavior is not specific to the data example, we replicate the simulation study of \citet{walz2024easyuq}.
In each iteration, we generate $n=5\,000$ predictor--outcome pairs with $X\sim\mathrm{Unif}(0,10)$, and
\[
Y \mid X \sim \mathrm{Gamma}\bigl(\text{shape}=\sqrt{X},\ \text{scale}=\sigma(X)\bigr),
\]
where $\sigma(X)=\min\{\max\{X,2\},8\}$.
We fit IDR on an 80\% training split and compute PIT values on the remaining 20\% test split. We repeat this procedure $N = 2\,000$ times to obtain 2 million out-of-sample PIT values in total.
As shown in the right panel of Figure~\ref{fig:pit_binning}, the resulting PIT values exhibit the same qualitative concentration near simple rational fractions as in the weather forecasting example.

Figure \ref{fig:thomae_like} shows empirical frequencies of individual PIT values. Here, we observe a distinct pattern resembling Thomae's function, with additional clusters at 0 and 1. The same pattern emerges in the weather application (not shown). The effect becomes clearly visible only when multiple IDR fits are used for prediction (in the simulation, one for each of the $N$ repeats; in the application, one for each grid point). With a single IDR fit, the clustering is noisier due to randomness in the training set and individual fits.

\section{THEORETICAL CONSIDERATIONS}

The spiking behavior can be traced to the pool block structure of isotonic regression.
In the univariate predictor case, EasyUQ estimates the conditional CDF
$F_x(y)=P(Y\le y\mid X=x)$ by fitting, for each threshold $y$, an isotonic regression for the binary exceedance response
$\mathbbm{1}\{y_i\le y\}$ against the predictor values $x_i$ \citep{henzi2021idr,walz2024easyuq}.
For fixed $y$, isotonic regression pools the training points based on the predictor values $x_i$ according to the PAV (pool-adjacent-violators) algorithm and fits a single value within each pool.
If a new predictor value $x$ is within the interval $[x_k,x_l]$ spanned by the predictor values in a pool $B = \{k, k+1,\dots,l\}$ (assuming that the training set is ordered according to the $x_i$) of size $\vert B\vert = l - k + 1$, the fitted value equals the pool average,
\[
\widehat F_x(y)=\frac{1}{|B|}\sum_{i\in B} \mathbbm{1}\{y_i\le y\},
\]
and therefore any such value is a rational number of the form $r/\vert B\vert$.
EasyUQ assembles these threshold-wise fits into a discrete conditional CDF $y\mapsto \widehat F_x(y)$ that is a step function in $y$ \citep{walz2024easyuq}.

\begin{figure}[t]
  \centering
  \includegraphics[width=0.925\linewidth]{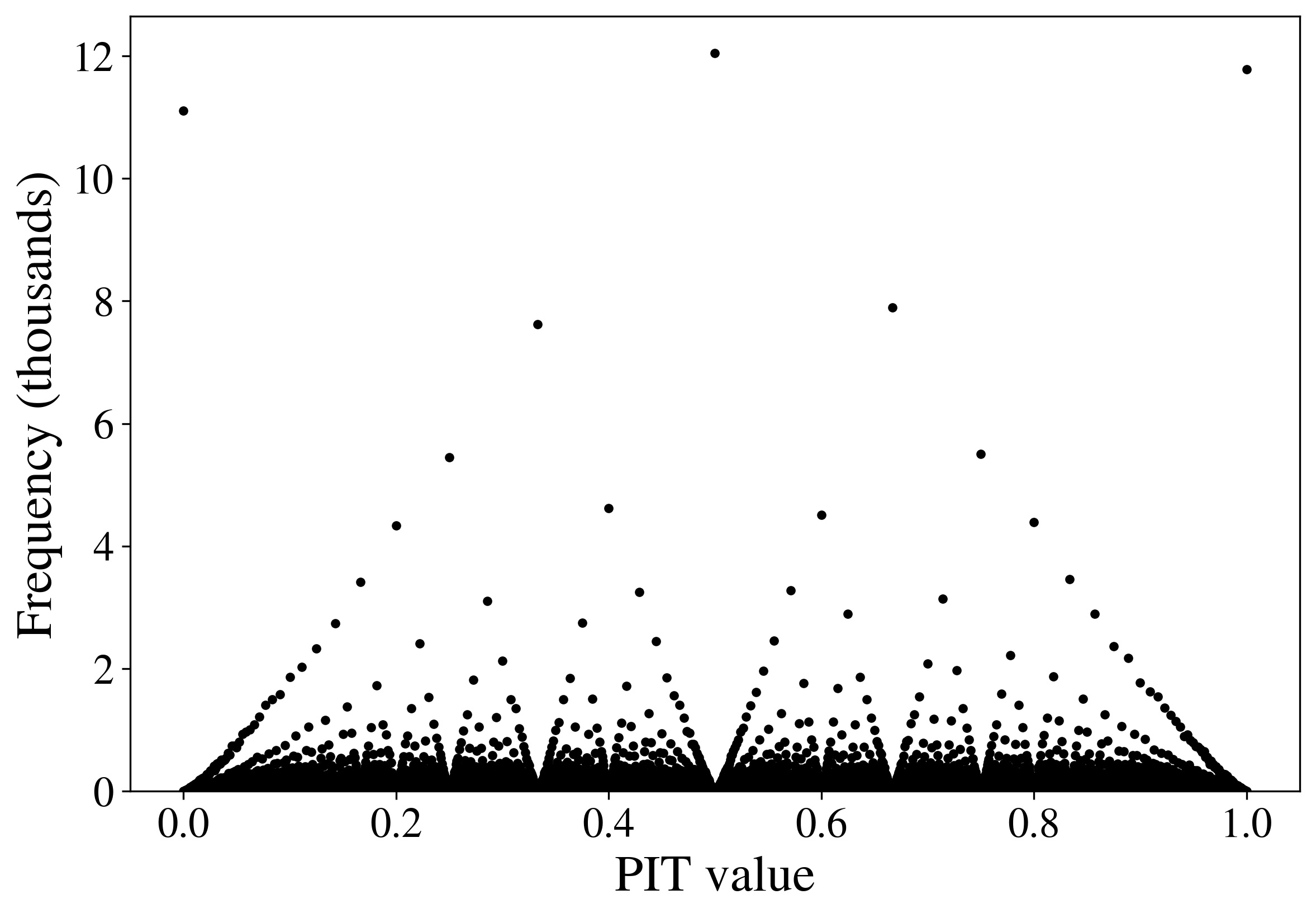}
  \caption{Empirical frequencies of unique PIT values under standard EasyUQ in the simulation study (test set). The distribution exhibits pronounced spikes at simple rational fractions (e.g., $1/2$, $1/3$, $2/3$, $1/4$, $3/4$), yielding a Thomae's function-like pattern.}
  \label{fig:thomae_like}
\end{figure}

For a new test pair $(x,y)$, the randomized PIT coincides with the ordinary PIT $\widehat F_x(y)$ with probability one under continuity of $Y$, since the probability of $y$ matching a jump point of the CDF is zero. 
Hence, when $x$ falls within a pool $B$ of the isotonic regression at threshold $y$, the PIT value $\widehat F_x(y)$ is a fraction of the form $r/\vert B\vert$, as above. We argue that the frequency of a reduced fraction $p/q$ as PIT value is proportional to the ratio of counting numbers divisible by the denominator $q$, under the assumption that the remainder of $\vert B\vert$ modulo $q$ is roughly uniformly distributed and $r$ given $\vert B\vert$ is approximately uniform across $0,1,\dots,\vert B\vert$.
Thus, when considering the test pair $(x,y)$ and training sample $(x_1,y_1),\dots,(x_n,y_n)$ to be iid random pairs with the same distribution to capture the randomness in both IDR fits and test samples, we obtain (for small, coprime positive integers $p,q$, assuming for simplicity that $n$ is divisible by $q$)
\begin{align*}
P\big(\tfrac{r}{\vert B \vert} = \tfrac{p}{q}\big) &= \sum_{i = 1}^{n/q} P(\vert B\vert = iq)\, P(r = ip \mid \vert B \vert = iq) \\ &\approx \sum_{i = 1}^{n/q} P(\vert B\vert = iq)\, \tfrac{1}{iq+1} =: S_0(q).
\end{align*}
The last sum is a partial sum of
\begin{align*}
S
&= \sum_{i=1}^n P(|B|=i)\,\frac{1}{i+1} \\
&= \sum_{j=0}^{q-1} \underbrace{\sum_{i=1}^{n/q} P(|B|=iq-j)\,\frac{1}{iq-j+1}}_{=:S_j(q)} .
\end{align*}
As long as the partial sums are all approximately equal, i.e., $S_j(q) \approx S_0(q)$ for all $j$, which should be the case as long as the distribution of $\vert B\vert$ is sufficiently broad, we have $S \approx qS_0(q)$, and thus empirical frequencies will be roughly proportional to $\frac1q$, explaining the Thomae's function-like pattern observed in Figure~\ref{fig:thomae_like}.

Note that there is a small (about 1.1\% in the simulation) chance of falling in between pools, in which case the PIT value is not a simple fraction (due to interpolation between fitted values). Furthermore, PIT values of zero or one may arise with any pool size or when a new observation is outside the range of training observations ($y < \widetilde{y}_1$ or $y > \widetilde{y}_k$).

\section{SMOOTHING}
\label{sec:smooth}

\begin{figure*}[t]
  \centering
  \includegraphics[width=\linewidth]{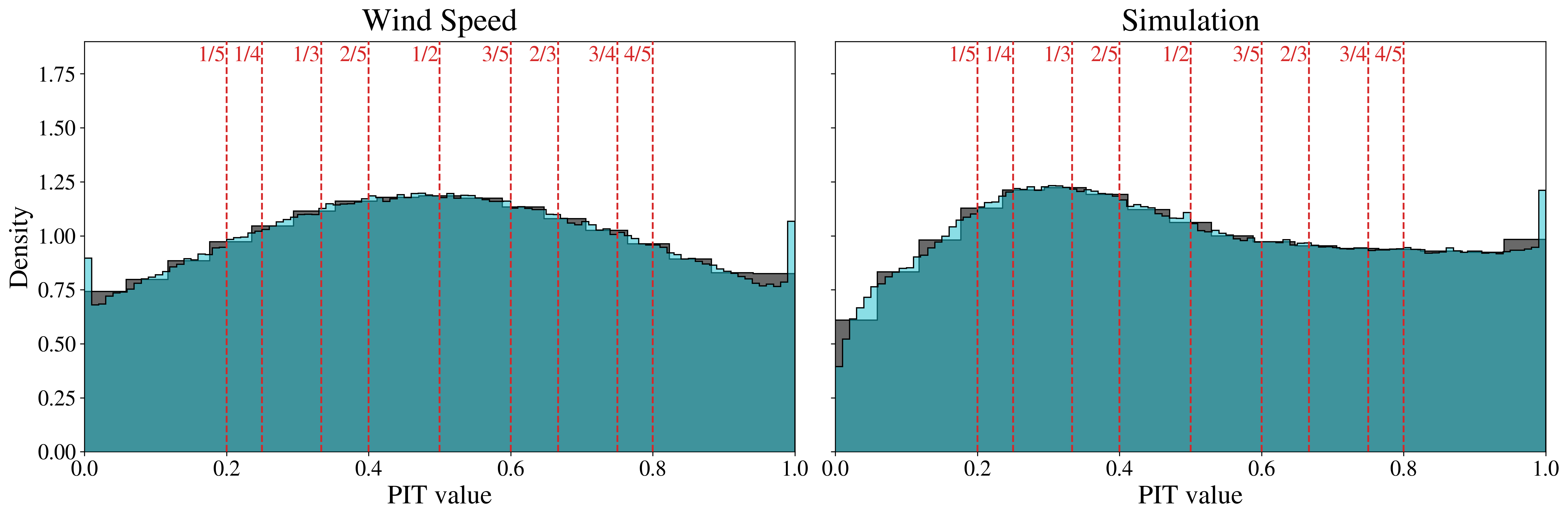}
  \caption{10-m wind speed (left) and simulation (right) out-of-sample PIT histograms for Smooth EasyUQ forecasts with coarse (17-bin; gray) and fine (100-bin; cyan) binning. 
  }
  \label{fig:pit_smooth}
\end{figure*}

\citet{walz2024easyuq} propose the Smooth EasyUQ approach, which supplements IDR with kernel smoothing and yields continuous predictive distributions. 
We use a Gaussian kernel leading to the smooth predictive CDF
\[
\widetilde F_x(y) = \sum_{j = 1}^k w_j(x) \Phi \left( \frac{y - \widetilde y_j}{h} \right),
\]
where $\Phi$ denotes the standard Gaussian CDF, $\widetilde y_j$ denotes the $j$th unique training observation, and the mixture weights correspond to the jumps in the CDF of the IDR solution, $w_j(x) = \widehat F_{x}(\widetilde y_j) - \widehat F_x(\widetilde y_{j - 1})$ where $\widehat F_x(\widetilde y_{0}):= 0$.
The bandwidth $h$ is selected using the one-fit log-score approach of \citet{walz2024easyuq}: after fitting IDR once on the training data, we choose $h$ to minimize the mean log score of the training observations under the smoothed model, using a leave-one-out style correction where the mixture weight $w_j(\cdot)$ corresponding to the left-out observation $y_i = \widetilde y_j$ is set to zero and the remaining weights are renormalized.
For the weather application, we tune $h$ independently on 250 randomly chosen grid points and use the median of the resulting values as a single global bandwidth applied across all grid points; we proceed analogously in the simulation study.

Empirically, the mean CRPS, a common measure of overall predictive performance \citep[e.g.,][]{GR07}, is almost unchanged under smoothing: 0.4397 (IDR) vs.\ 0.4395 (smoothed) for the real data, and 4.1591 vs.\ 4.1654 in the simulation. Figure~\ref{fig:pit_smooth} shows that the smoothing approach effectively mitigates spikes in the PIT histogram.
However, smoothing yields some clear deviations from uniformity in the PIT histogram that were not present with the original IDR predictions (cf.\ Figure \ref{fig:pit_binning}).

\section{CONCLUSIONS}

We have documented a peculiar pattern in the PIT values of IDR predictions, which can be observed in both simulated and real data. While this observation does not contradict existing theory on the calibration of IDR, the observed pattern is in contrast to various previous applications of IDR, where well-calibrated forecasts were reported. 
While deviations of the PIT values from uniformity are to be expected because of the discreteness of IDR predictions, we were surprised by the striking Thomae-like pattern in empirical frequencies. We believe this pattern to be of a theoretical interest in its own right, hinting at interesting regularities in the distribution of the PIT values obtained from IDR. 

Clearly, some smoothing is required to obtain appropriate continuous CDF predictions.
\citet{walz2024easyuq} suggest kernel smoothing as in their Smooth EasyUQ approach as a remedy that should be preferred for continuous outcomes. Yet, many practitioners appear to still use the standard (unsmoothed) IDR. Smooth EasyUQ requires some tuning via the selection of a bandwidth (and kernel) and appears to introduce some deviations from probabilistic calibration of its own that cannot be traced to the underlying IDR solution. 
We also experimented with evaluating PIT values after replacing the piecewise-constant predictive CDF with a simple piecewise-linear interpolation, which effectively removed the Thomae-like pattern. Except for accumulation points at zero and one due to missing tail extrapolation, this approach produced continuous PIT values resulting in roughly uniform looking PIT histograms in our simulation. It remains an open problem how to smooth IDR in a way that produces well-calibrated predictions. 

\clearpage 
\subsubsection*{Acknowledgements}

We thank Tilmann Gneiting, Kristof Kraus, and Johanna Ziegel for insightful comments and fruitful discussion. TB gratefully acknowledges support by the German Weather Service (Deutscher Wetterdienst) through the SPARC-ML project within the extramural research program, funding reference number 4823EMF01. JR gratefully acknowledges support from the Deutsche Forschungsgemeinschaft (DFG, German Research Foundation) via project number 502572912. AJ is grateful for the generous support of the Klaus Tschira Foundation. SL gratefully acknowledges support by the Vector Stiftung through the Young Investigator Group ``Artificial Intelligence for Probabilistic Weather Forecasting''.

\subsubsection*{Code and Data Availability}
Code for replicating the simulation study and the empirical analysis is available at \url{https://github.com/tobiasbiegert/idr-calibration}. WeatherBench 2 data are freely accessible; see \cite{rasp2024weatherbench}.

\bibliography{references}

\end{document}